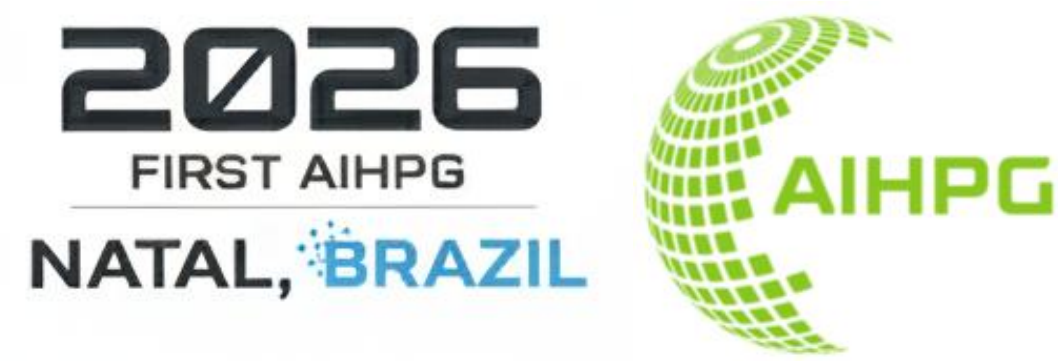

# Synthetic Well Log Generation with Preserved Multivariate Correlations and Vertical Facies Stacking Patterns

**Fonseca, Josue[1] and Saraiva, Marcus[1]**
*[1]Petrobras: Av Henrique Valadares, 28, Torre B, 6º andar, 20231-030, Centro, Rio de Janeiro, Brazil*
*Contact email: josuefonseca@petrobras.com.br*

**ABSTRACT.** We present a novel procedure for generating synthetic well logs that simultaneously preserves multivariate correlations among petrophysical properties (Density, P-Sonic, S-Sonic) and vertical stacking patterns of electrofacies. The methodology integrates Markov chain models, autoencoder-based dimensionality reduction, and Markov chain Monte Carlo (MCMC) sampling in latent space. Application to a real turbidite reservoir dataset demonstrates that the framework successfully sustains fundamental rock physics relationships and generates geologically realistic vertical heterogeneity consistent with actual well log measurements. This technique addresses critical data scarcity in machine learning applications for seismic interpretation while enabling credible synthetic seismogram generation for scenario testing and uncertainty quantification in petroleum exploration and field development.

## INTRODUCTION

Synthetic seismic modeling plays a crucial role in modern geophysical workflows, enabling geoscientists to understand seismic responses across diverse geological scenarios. The generation of realistic synthetic seismograms depends on accurate petrophysical property logs, particularly bulk density (mnemonic used during well logging: RHOB), compressional sonic (mnemonic: DT), and shear sonic (mnemonic: DTS), which are intrinsically linked through rock physics relationships (Avseth et al., 2005).

Conventional simulation approaches exhibit two critical limitations. First, they fail to capture complex multivariate correlations between rock physical properties—density, compressional velocity, and shear velocity exhibit non-linear relationships governed by lithology, porosity, and fluid content (Greenberg and Castagna, 1992). Second, they neglect vertical stacking patterns of facies observed in real geological successions (Doveton and Cable, 1986), resulting in geologically implausible models. These limitations are particularly problematic for machine learning applications requiring large volumes of annotated training data, as real well data with comprehensive log suites remain scarce and expensive.

This work proposes an innovative methodology integrating Markov chain models, autoencoder-based dimensionality reduction, and MCMC sampling to generate synthetic well logs that simultaneously preserve multivariate correlations between RHOB, DT, and DTS while honoring vertical stacking patterns of electrofacies. We validate the approach using a real turbidite reservoir dataset, enabling both realistic forward seismic modeling and systematic augmentation of datasets for training machine learning models.

## METHODOLOGY

The methodology integrates statistical modeling and machine learning in a four-stage workflow:

Stage 1: Electrofacies Definition and Vertical Stacking Modeling

Electrofacies characterization establishes the geological framework, representing distinct petrophysical responses corresponding to specific lithological characteristics (Busch et al., 1987). We analyze vertical succession patterns to construct an empirical transition probability matrix T, where $T_{ij}$ represents the probability of transitioning from facies i to facies j vertically (Carle and Fogg, 1996). This matrix feeds a discrete-step Markov chain model that simulates vertical facies sequences, ensuring synthetic profiles exhibit realistic organization consistent with observed stratigraphic patterns.

Stage 2: Dimensionality Reduction via Autoencoder

We employ a regression autoencoder with Multi-Layer Perceptron (MLP) architecture to compress the three-dimensional property space (Density, P-Sonic, S-Sonic) into a two-dimensional latent representation (Hinton and Salakhutdinov, 2006). Properties are standardized prior to training. The autoencoder minimizes mean squared error between original and reconstructed dataset properties, forcing the latent space to capture essential non-linear correlations among properties (Bank et al., 2021). This dimensionality reduction provides computational efficiency for subsequent sampling while preserving complex interdependencies observed in real rock physics.

Stage 3: Multivariate Sampling via Markov Chain Monte Carlo

We generate synthetic samples in latent space using the Metropolis-Hastings (MH) algorithm, sampling from empirical distributions estimated by kernel density estimation for each electrofacies (Bishop, 2006). The trained decoder network maps latent coordinates back to original property space, and with some transformations yielding synthetic (Density, P-Velocity, S-Velocity) triplets. This process ensures generated property combinations respect multivariate correlations and rock physics constraints observed in real data.

Stage 4: Synthetic Well Log Assembly

Complete synthetic logs are assembled by: (a) generating vertical facies profiles using the Markov chain model at discrete depth intervals, (b) retrieving MCMC-generated property samples corresponding to assigned facies at each depth, and (c) concatenating values vertically to construct continuous logs. This yields synthetic logs honoring both vertical facies transition probabilities and petrophysical property correlations consistent with assigned facies.

## RESULTS AND DISCUSSION

We applied the methodology to a turbidite reservoir dataset comprising four wells within a specific stratigraphic interval.

Electrofacies Characterization and Vertical Transition Patterns – Stage 1

Figure 1a displays electrofacies distribution across four wells. Petrophysical clustering revealed four distinct electrofacies representing characteristic turbidite lithological variability. Figure 1b presents the empirical transition probability matrix capturing stratigraphic trends that govern vertical organization of synthetic facies profiles

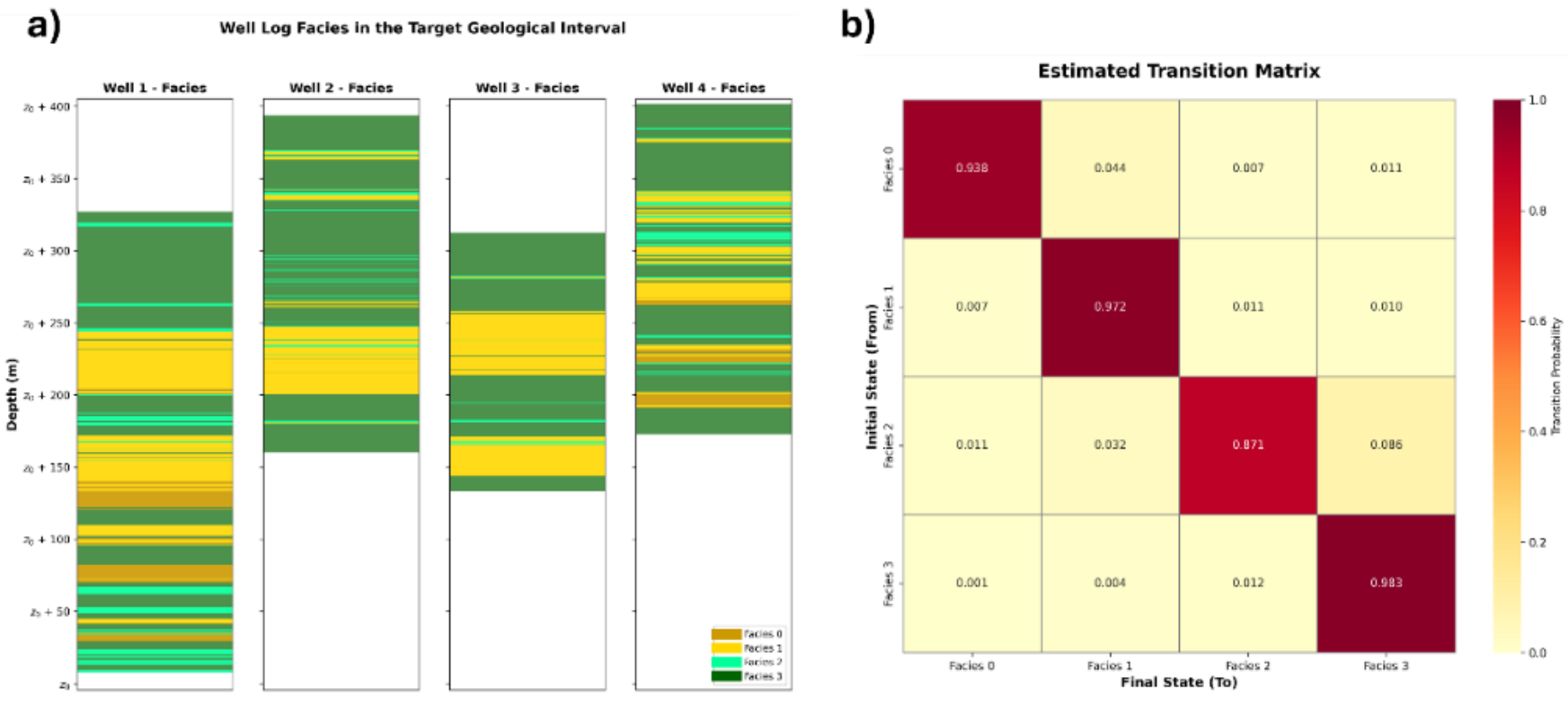

Figure 1: Real dataset with four wells with: (a) facies classification and (b) transition matrix T of vertical facies changes from bottom to top represented by a heatmap.

Dimensionality Reduction and Latent Space Representation – Stage 2

Figure 2 illustrates dimensionality reduction achieved by the autoencoder. The latent space preserves essential data structure despite dimensionality reduction (3D to 2D), enhancing facies separation and manifold geometry encoding property correlations. This demonstrates that dominant variability in the three properties can be captured by two latent dimensions, suggesting strong correlations driven by common geological controls such as lithology and porosity (Castagna et al., 1985).

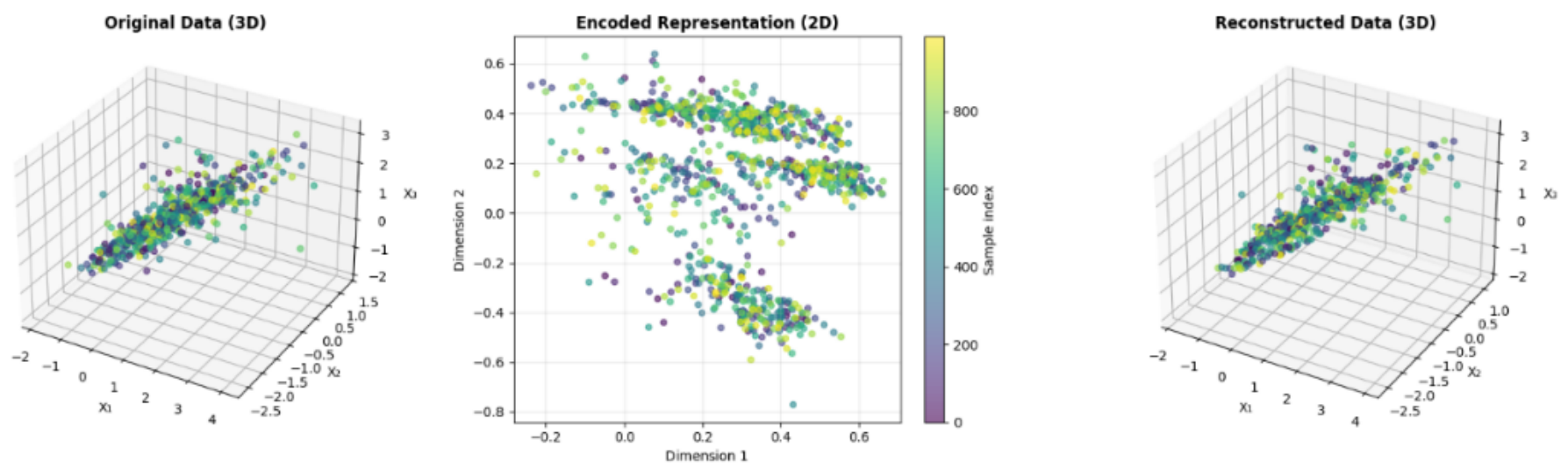

Figure 2: Three plots representing the input, bottleneck. and output of the autoencoder with the well-log dataset.

Multivariate Sampling and Property Correlation Preservation – Stage 3

Figure 3 demonstrates MCMC sampling effectiveness in reproducing statistical characteristics of encoded real log data. The close agreement between target distribution and MCMC samples confirms successful latent space exploration, generating synthetic representations statistically consistent with observed data.

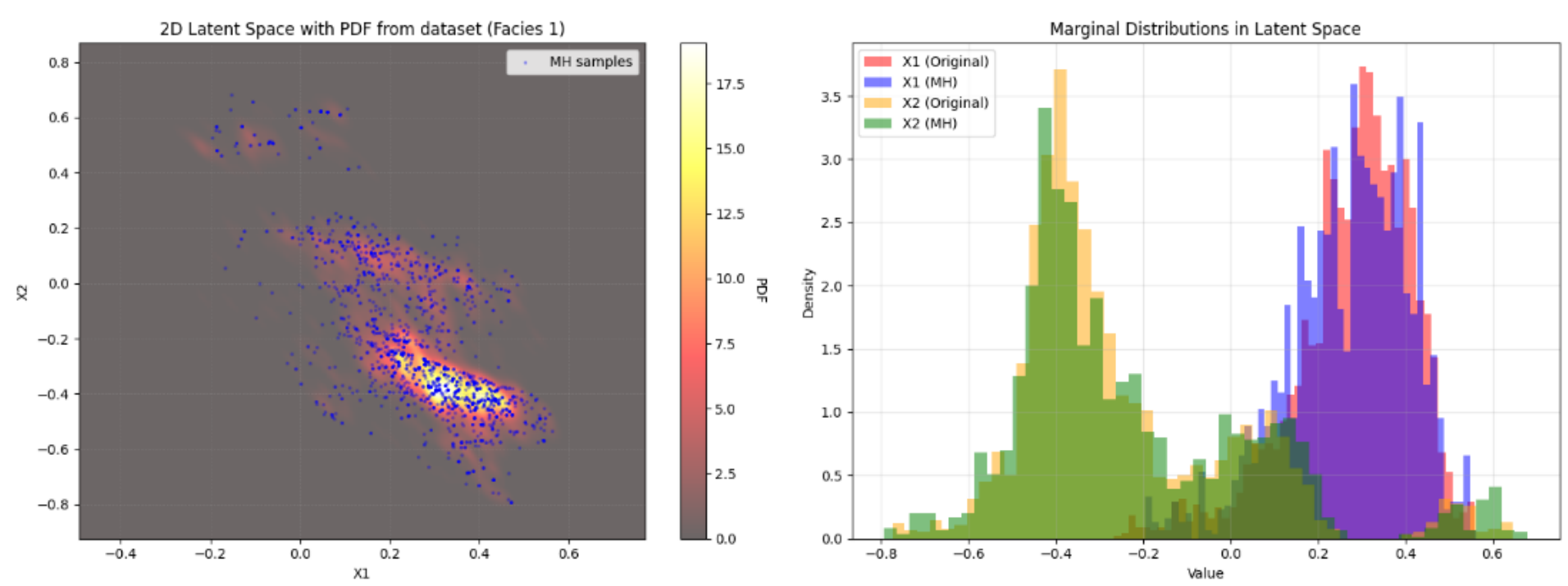

Figure 3: Using the latent space to perform MCMC sampling through MH algorithm.

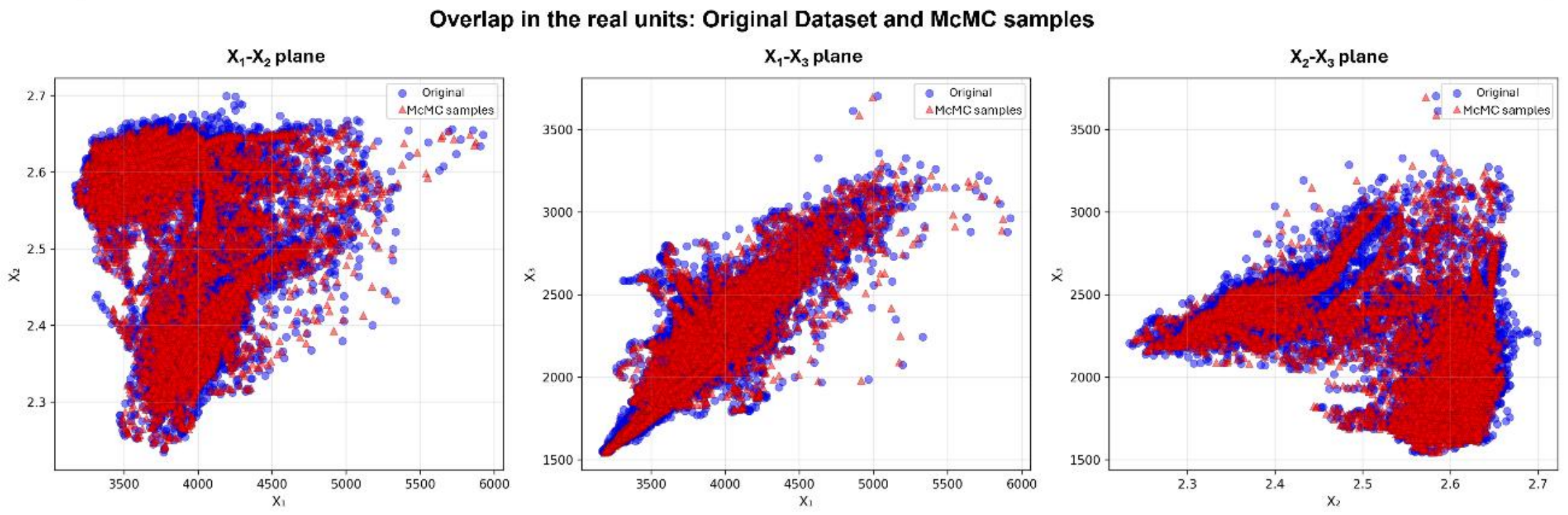

Figure 4: Original dataset (blue dots) versus MCMC samples (red dots) in the real units of the rock properties.

Figure 4 presents pairwise scatter plots comparing real well log measurements with synthetic samples after decoding. The synthetic samples closely replicate correlation structures, trends, and scatter patterns observed in real data, critical for generating physically plausible property combinations respecting rock physics constraints (Mavko et al., 2009).

Synthetic Well Log Generation and Geological Realism – Stage 4

Figure 5 presents the generated ensemble of synthetic well logs spanning 250 meters of depth, showing simulated electrofacies profiles alongside corresponding density, P-Velocity ($V_p$), and S-Velocity ($V_s$) curves. The synthetic logs exhibit realistic vertical facies organization, property-facies consistency, high-frequency variability, and appropriate inter-realization variability.

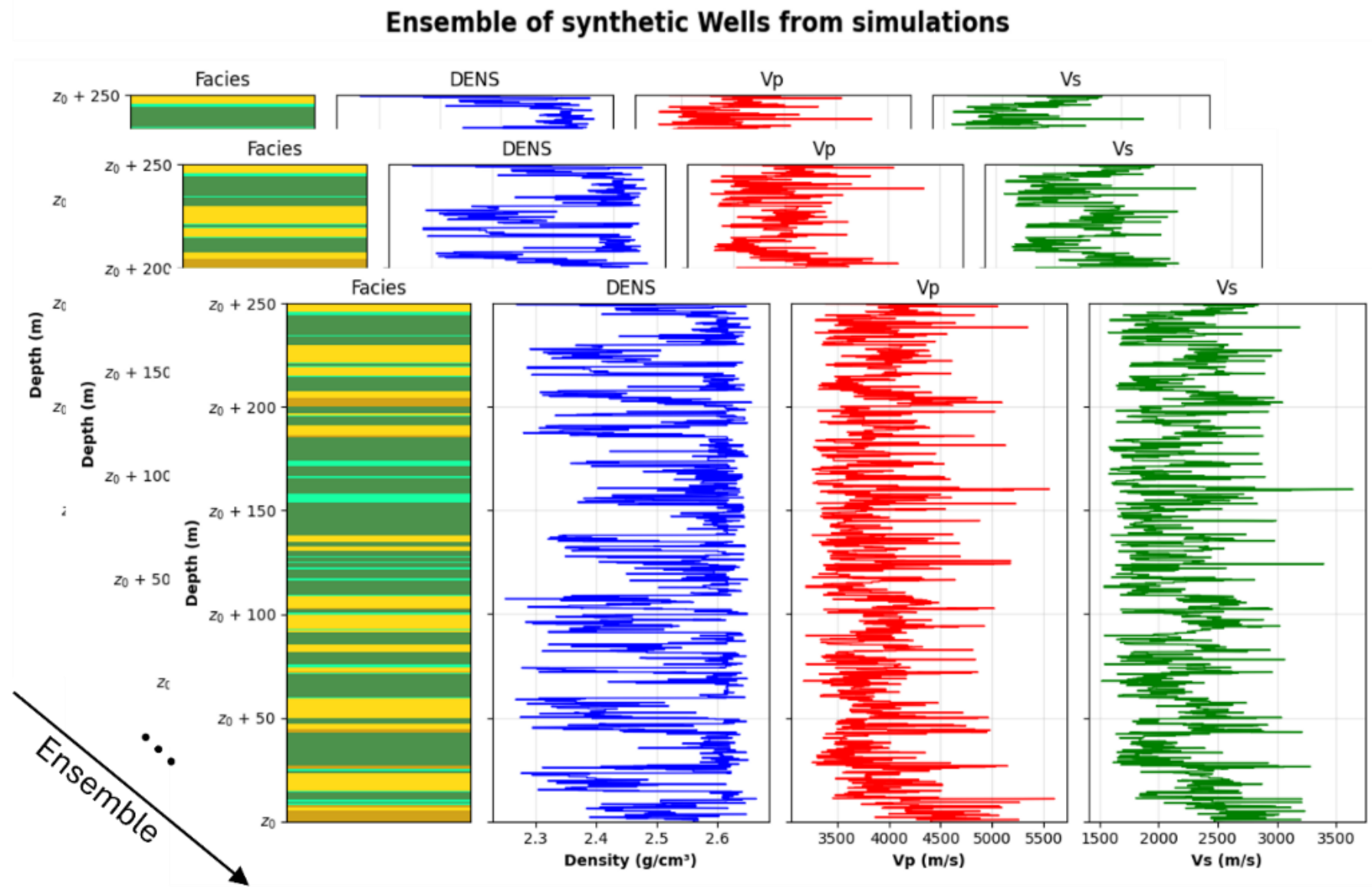

Figure 5: Generated ensemble of facies and well-log suite for further geophysical modeling.

For seismic forward modeling, preserved rock physics relationships and vertical facies organization ensure synthetic seismograms exhibit realistic amplitude variations, frequency content, and stratigraphic patterns. For machine learning applications, the ability to generate large volumes of synthetic training data with known ground truth addresses critical bottlenecks in developing supervised learning models, with geological realism ensuring models learn meaningful relationships between seismic responses and subsurface properties.

## CONCLUSIONS

This work presented an innovative methodology for generating synthetic well logs that preserves multivariate correlations among petrophysical properties (Density, P-Sonic, S-Sonic) and vertical stacking patterns of electrofacies. By integrating Markov chain models, autoencoder-based dimensionality reduction, and MCMC sampling in latent space, the framework produces geologically realistic synthetic logs at scale. Application to a real turbidite reservoir demonstrated effectiveness in maintaining rock physics relationships and generating appropriate vertical heterogeneity consistent with actual well log measurements.

The proposed framework has significant implications for both seismic forward modeling and machine learning applications. The achieved geological realism enables generation of credible synthetic seismograms for scenario testing and uncertainty quantification, while scalability addresses data scarcity limiting development of automated interpretation tools. This methodology supports adoption of artificial intelligence technologies in seismic interpretation while maintaining geological fidelity essential for petroleum exploration and field development. Future extensions could incorporate additional petrophysical properties and adapt the framework to other depositional environments.

## ACKNOWLEDGMENTS

The authors thank Petrobras for its support.